\documentclass[a4paper,fleqn]{cas-dc}

\usepackage{subfig}
\usepackage{threeparttable}
\usepackage[numbers]{natbib}
\setcitestyle{square}

\def\tsc#1{\csdef{#1}{\textsc{\lowercase{#1}}\xspace}}
\tsc{WGM}
\tsc{QE}

\begin{document}
\let\WriteBookmarks\relax
\def\floatpagepagefraction{1}
\def\textpagefraction{.001}

\shorttitle{}    

\shortauthors{Z. Zhou et~al.}  

\title [mode = title]{Mixer-Informer-Based Two-Stage Transfer Learning for Long-Sequence Load Forecasting in Newly Constructed Electric Vehicle Charging Stations}

%

\author[]{Zhenhua Zhou}[type=editor,
                        orcid=0009-0005-1692-6208]

\fnmark[a]

\ead{zhenhua.zhou@connect.polyu.hk}

\author[]{Bozhen Jiang}

\fnmark[a]
\ead{bozhen.jiang@connect.polyu.hk}

\author[]{Qin Wang}

\cormark[1]

\fnmark[a]
\ead{qin-ee.wang@polyu.edu.hk}
\cortext[1]{Corresponding author}

\credit{}

\affiliation[1]{organization={Department of Electrical and Electronic Engineering},
            addressline={The Hong Kong Polytechnic University}, 
            city={Hung Hom},
            postcode={}, 
            state={Kowloon},
            country={Hong Kong}}


\begin{abstract}
The rapid rise in electric vehicle (EV) adoption demands precise charging station load forecasting, challenged by long-sequence temporal dependencies and limited data in new facilities. This study proposes MIK-TST, a novel two-stage transfer learning framework integrating Mixer, Informer, and Kolmogorov-Arnold Networks (KAN). The Mixer fuses multi-source features, Informer captures long-range dependencies via ProbSparse attention, and KAN enhances nonlinear modeling with learnable activation functions. Pre-trained on extensive data and fine-tuned on limited target data, MIK-TST achieves 4\% and 8\% reductions in MAE and MSE, respectively, outperforming baselines on a dataset of 26 charging stations in Boulder, USA. This scalable solution enhances smart grid efficiency and supports sustainable EV infrastructure expansion.
\end{abstract}



\begin{keywords}
Mixer, Informer, KAN, Electric Vehicle, Charging Station Forecasting
\end{keywords}

\maketitle
\section{Introduction}

The widespread adoption of electric vehicles (EVs) marks a pivotal shift toward sustainable transportation, with global sales almost 14 million units in 2023 alone~\cite{iea2024}. The rapid growth in Electric Vehicle (EV) ownership has spurred a corresponding expansion of charging infrastructure, thereby exacerbating the demand for accurate load forecasting in EV charging stations\cite{zhu2019electric}. Precise charging load prediction is critical for smart grid management, as it ensures grid stability, optimizes energy distribution, and facilitates the seamless integration of renewable energy sources in modern urban environments. However, EV charging load forecasting presents inherent challenges, primarily due to the long-sequence dependencies inherent in the data, the influence of multi-dimensional external factors, and the scarcity of historical data, particularly in newly constructed stations\cite{wang2023transfer,wang2025electric}. These challenges limit the effectiveness of traditional forecasting methods, which necessitates the development of more advanced models to enhance prediction accuracy.
	
Electric vehicle charging demand exhibits intricate temporal patterns characterized by multiple periodic components, including daily, weekly, and seasonal cycles. In urban environments, this demand is further shaped by various external factors such as weather conditions, public holidays, and local traffic dynamics \cite{al2019review}. The interaction of these multidimensional influences leads to a highly non-uniform distribution of charging loads, posing challenges for traditional statistical models like ARIMA and exponential smoothing \cite{shumway2000time}, which rely on assumptions of linearity and stationarity. Due to these inherent limitations, deep learning approaches—particularly Transformer-based architectures—have gained prominence in time-series forecasting. The Informer model \cite{zhou2021informer}, with its ProbSparse attention mechanism, demonstrates superior performance in long-sequence modeling while significantly reducing the computational complexity associated with traditional self-attention mechanisms.
	
In recent years, researchers have proposed various advanced deep learning architectures for long-sequence time-series forecasting, aiming to improve prediction accuracy across diverse application scenarios. For instance, PatchTST \cite{nie2022time} ecaptures local temporal features by decomposing time-series data into non-overlapping sub-sequences (patches) and aggregating them using the Transformer architecture. However, the fixed-size patches of this model may fail to adapt to the complexity of multi-scale temporal dependencies in charging load data. Autoformer \cite{wu2021autoformer} enhances periodic pattern modeling through decomposition modules and auto-correlation mechanisms but suffers from high computational complexity, which can hinder its scalability for large datasets. Besides, DLinear \cite{chen2022transformers}, which employs a stacked linear layer approach, reduces computational overhead while maintaining high prediction accuracy. Nevertheless, its linear architecture may struggle to effectively capture the prevalent non-linear dynamics in charging load data. In addition, Crossformer \cite{zhang2023crossformer} introduces a cross-dimensional attention mechanism to model dependencies across multiple time-series variables. However, it fails to address the issue of data sparsity in newly established charging stations, which limits its effectiveness in transfer learning scenarios. Meanwhile, FreTS \cite{yi2023frequency} leverages Fourier transforms to extract frequency-domain features, improving efficiency by filtering noise and learning key spectral components \cite{forootani2023transfer}. However, frequency-domain representations alone may not fully preserve the temporal dynamics crucial for charging load forecasting.

	While models like PatchTST, Autoformer, DLinear, Crossformer, and FreTS have significantly advanced long-sequence time-series forecasting, they often lack effective transfer learning strategies, and thereby making them less suitable for scenarios with data scarcity, such as newly constructed charging stations. As a result, their generalization ability in new environments remains limited. To address these challenges, this paper proposes MIK-TST, a novel framework that integrates the strengths of Mixer \cite{tolstikhin2021mlp}, Informer \cite{zhou2021informer}, and KAN \cite{liu2024kan}, coupled with a two-stage transfer learning strategy. Specifically, the Mixer architecture enables effective fusion of multi-source heterogeneous features, including temporal and channel dimensions, providing more comprehensive input information. The ProbSparse attention mechanism of the Informer module effectively captures the long-term dependencies and periodic patterns of charging load data. Meanwhile, the KAN module captures the complex nonlinear relationships with greater precision while improving interpretability through its learnable activation functions. Furthermore, the two-stage transfer learning strategy further enhances the model's adaptability and prediction accuracy in new scenarios by first pre-training on multiple established charging stations and subsequently fine-tuning on data-scarce target stations. The main contributions of this work are as follows:

	\begin{enumerate}
		\item \textbf{MIK Hybrid Architecture:} This paper proposes a novel MIK hybrid architecture integrating Mixer, Informer, and KAN modules. Mixer effectively fuses multi-source temporal and channel features; Informer efficiently captures long-term temporal dependencies; KAN improves non-linear approximation capability and model interpretability.
		
		\item \textbf{Two-Stage Transfer Learning Strategy:} A two-stage transfer learning framework is introduced to mitigate the reliance on extensive station-specific data, facilitating rapid adaptation and deployment in new charging stations. This approach involves pre-training on large, multi-station datasets, followed by fine-tuning on data-scarce target stations, thereby significantly reducing data dependency.
		
		\item \textbf{Enhanced Prediction Performance:} Empirical evaluations conducted on a dataset of 26 charging stations in Boulder, USA, demonstrate that the proposed model outperforms the baseline methods, achieving improvements of 4\% to 8\% in forecasting accuracy metrics
		
		\item \textbf{Practical Significance for Smart City Applications:} The proposed method provides a scalable and robust solution for smart grid management and urban energy planning, ensuring efficient energy distribution and supporting the sustainable expansion of electric vehicle infrastructure.
	\end{enumerate}

    The remainder of this paper is structured as follows: Section Preliminaries outlines the problem formulation and reviews related work; Section Methodology details the MIK-TST framework; Section Experiments presents experimental results addressing three research questions; and Conclusion Section summarizes findings and future directions.
	
	\section{PRELIMINARIES}
        \subsection{Problem Formulation}    
    \textbf{Input}: Multi-dimensional features, such as historical charging power, timestamps, charging session logs, or other relevant station-level indicators (e.g., day-of-week flags). Prior to model training, the raw input data  is often preprocessed by standardization to ensure consistent scaling across features:
	\begin{equation}
	X_{std} = \frac{x_i - u}{\theta}
	\end{equation}
    
    \textbf{Output}: Predictions of charging load for future time horizons (e.g., the upcoming 24 hours, days, or weeks). The proposed framework leverages the Mixer module to effectively fuse multiple feature dimensions, the Informer module to efficiently capture long-sequence temporal dependencies, and the KAN module to model complex non-linear relationships. This integrated approach aims to handle both intricate temporal patterns and station-specific variations in electric vehicle charging load forecasting.

	\subsection{Related Studies on EV Charging Load Forecasting:}
    The increasing adoption of electric vehicles has spurred significant research into charging load forecasting, a critical component of intelligent energy management systems. For instance, Zhu et al. \cite{zhu2019electric}  conducted a comparative study of deep learning approaches for EV charging load forecasting, utilizing Long Short-Term Memory (LSTM) networks to capture temporal dependencies in charging data from urban stations. Their work demonstrated promising results for short-term predictions but struggled with long-sequence forecasting due to vanishing gradient issues. Similarly, Wang et al. \cite{wang2023transfer} proposed a transfer learning method based on deep reinforcement learning to optimize charging strategies across multiple stations. Their approach successfully adapted pre-trained models to new stations but relied heavily on extensive source domain data, limiting its scalability in data-scarce scenarios. More recently, Wang et al. \cite{wang2025electric} explored the impact of weather on charging load forecasting, integrating meteorological data into a convolutional neural network (CNN) model. While effective in capturing short-term weather-induced fluctuations, their method lacked a robust mechanism for modeling long-term periodic patterns, such as weekly or seasonal trends.
    
    Other notable efforts include Al-Ogaili et al. \cite{al2019review}, who reviewed scheduling and clustering techniques for EV charging load prediction, emphasizing the role of user behavior patterns. Their findings highlighted the challenge of high-dimensional feature spaces but did not address the computational inefficiencies of processing long-sequence data. In contrast, advanced time-series models like PatchTST \cite{nie2022time}, Autoformer \cite{wu2021autoformer}, and Crossformer \cite{zhang2023crossformer} have shown superior performance in general long-sequence forecasting tasks. However, these models often lack tailored transfer learning strategies, making them less effective for newly constructed charging stations where historical data is limited. This gap underscores the need for a hybrid framework like MIK-TST, which combines feature fusion, long-range dependency modeling, and transfer learning to address the unique challenges of EV charging load forecasting in data-scarce environments.

\begin{figure*}[!t]
	\centering
	\includegraphics[scale=0.3]{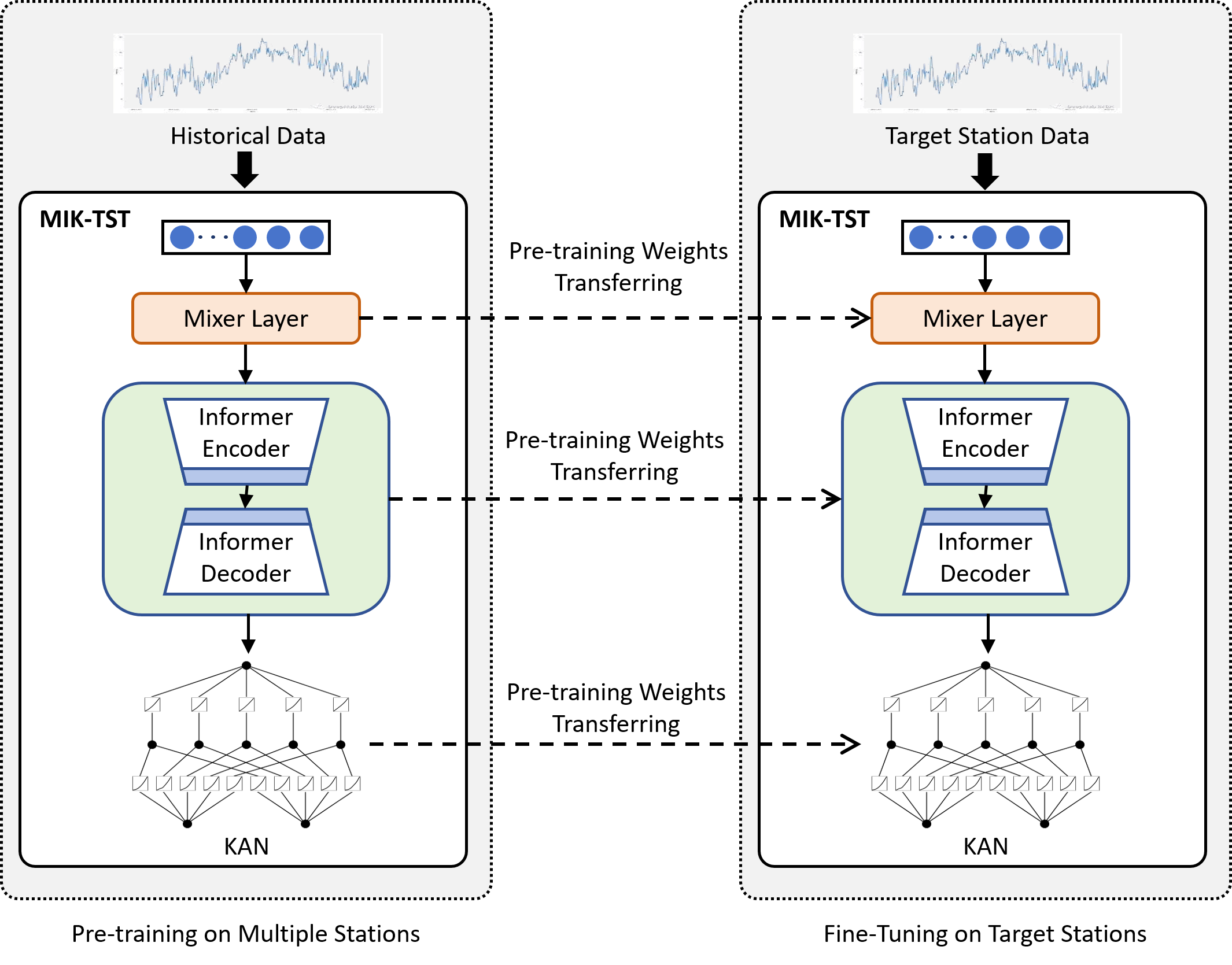}
	\caption{Overall architecture of the MIK-TST}
        \label{MIK-TST}
\end{figure*}

	\section{METHODOLOGY}
	The MIK-TST framework combines Mixer, Informer, and KAN within a two-stage transfer learning pipeline (Fig.~\ref{MIK-TST}). Input features are fused by Mixer, processed by Informer for temporal modeling, and refined by KAN for nonlinear mapping, with pre-training on source data followed by fine-tuning on target data. This section details each module and the transfer strategy.

	\subsection{Mixer Module}
	The Mixer module is a core component proposed in this study, designed to efficiently integrate information across both temporal and feature dimensions of input data, thereby capturing complex dependencies in multivariate time series.
	
	\begin{figure}[!t]
		\centering
		\includegraphics[scale=0.35]{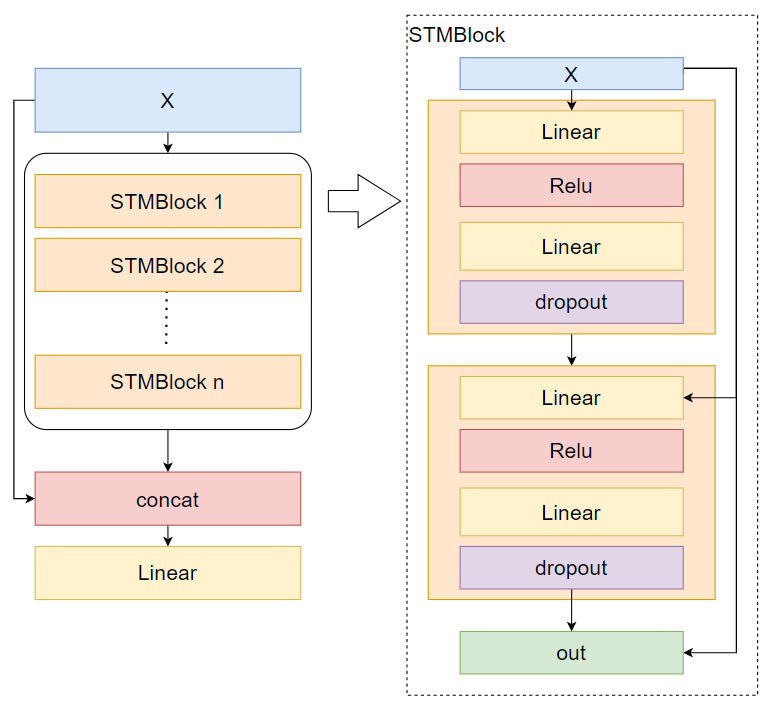}
		\caption{Overall architecture of the Mixer}
	\end{figure}
	
	\subsubsection{Network Structure}
	The basic structure of the Mixer module consists of two main mixing layers: the temporal mixing layer and the feature mixing layer.
	
	\noindent\textbf{Temporal Mixing Layer:}
	
	\begin{itemize}
		\item This layer first performs linear projection on the input sequence followed by a fully connected (or convolutional) operation, aggregating information across multiple time steps.
	\end{itemize}
	\begin{itemize}
		\item It is proposed to capture long-range dependencies in the time series, enabling the model to effectively utilize historical information from distant past observations.
	\end{itemize}

	\noindent\textbf{Feature/Channel Mixing Layer:}
	
	\begin{itemize}
		\item Following temporal mixing, the feature mixing layer applies a linear projection and fully connected operations across different feature channels.
	\end{itemize}
	\begin{itemize}
		\item This layer facilitates interactions between features, allowing the model to capture the interdependencies among various input features.
	\end{itemize}
	\begin{itemize}
		\item TBy leveraging this mechanism, the model effectively learns cross-feature relationships.
	\end{itemize}
	
	\noindent The specific calculation process is exhibited as follows:
	\begin{equation}
	X_{Mixer} = f_L(\mathcal{M}(X))
	\end{equation}
	\begin{equation}
	f_L(X) = WX+b
	\end{equation}
	where $x \in \mathbb{R}^n$, $y \in \mathbb{R}^m$, $W \in \mathbb{R}^{m \times n}$, and $b \in \mathbb{R}^m$. The function $\mathcal{M}$ consists of multiple $STM\_blocks$, which can be defined as follows:
	\begin{equation}
	ST\_blocks(X) = X + f_C(X + f_T(X))
	\end{equation}
	\begin{equation}
	f_T(X) = \mathcal{M}(f_L(\sigma_{\text{ReLU}}(f_L(X))))
	\end{equation}
	\begin{equation}
	f_C(X) = \mathcal{M}(f_L(\sigma_{\text{ReLU}}(f_L(X))))
	\end{equation}
	where $\sigma_{\text{ReLU}}$ is the activation function \cite{glorot2011deep}, $\mathcal{M}$ is an effective regularization technique \cite{hinton2012improving}.

	\subsubsection{Advantages of the Mixing Mechanism}
	The Mixer module enhances the model’s predictive performance by generating information-rich feature representations through its feature mixing operations \cite{baltruvsaitis2018multimodal}. Specifically, the core mechanism of temporal domain mixing lies in aggregating long-range dependencies within the time series. By utilizing fully connected layers, the model is enabled to learn the correlations between different time steps, thus accurately capturing the evolutionary trends of the time series. This is of paramount importance for predicting data with significant temporal dependencies, such as electric vehicle charging loads \cite{lai2018modeling}. Moreover, the Mixer module’s structural simplicity and flexibility allow seamless integration with other advanced architectures (e.g., Informer), making it highly adaptable to diverse time-series prediction tasks \cite{tolstikhin2021mlp}.

	\subsection{Informer Module}
	The Informer model is an efficient model optimized based on the Transformer architecture, specifically designed for long-sequence time-series forecasting. Its core innovation lies in the introduction of the ProbSparse attention mechanism, which effectively reduces the computational complexity of standard self-attention, thereby significantly enhancing the ability to process long-sequence data.
	
	\begin{figure}[h]
		\centering
		\includegraphics[scale=0.5]{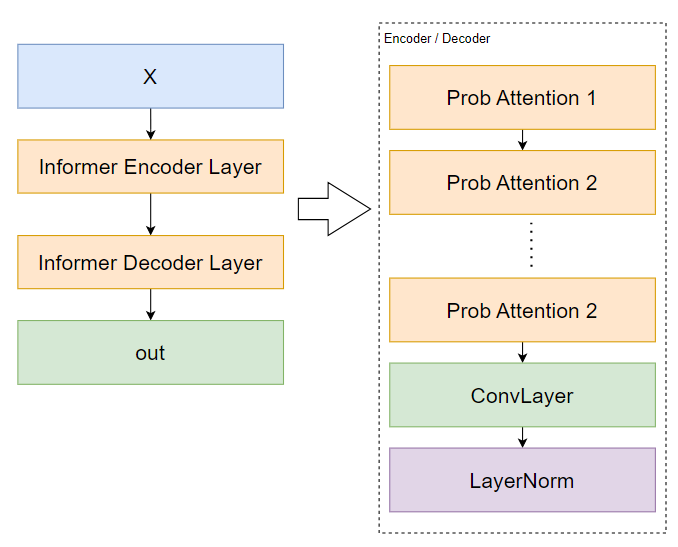}
		\caption{Overall architecture of the Informer}
	\end{figure}
	
	\subsubsection{Network Structure}
	
	The Informer model employs the ProbSparse attention mechanism, which can be mathematically expressed as:
	\begin{equation}
	\mathcal{A}(Q, K, V) = \sigma_{\text{softmax}}(QK^T / sqrt(d_k))V
	\end{equation}
	where Q (queries), K (keys), and V (values) are obtained by linear transformations of the input sequence and $d_k$ is the dimension of the key vectors. ProbSparse attention reduces unnecessary computation by probabilistically selecting key queries, thereby enabling effective processing of long sequences.
	The overall architecture of the Informer model follows the classic encoder-decoder design:
	\begin{itemize}
		\item \textbf{Encoder:} Composed of multiple stacked attention layers that progressively extract high-level feature representations from the input sequence. By employing ProbSparse attention, the encoder efficiently captures long-range dependencies while maintaining computational efficiency.
	\end{itemize}
	\begin{itemize}
		\item \textbf{Decoder:} Generates long-horizon predictions using the encoder’s outputs and, in some cases, partial input sequences. The decoder integrates ProbSparse attention with a generative decoding strategy, enabling accurate forecasting of future time steps.
	\end{itemize}
	
	\noindent The specific calculation formula is shown as follows:
	\begin{equation}
	X_{Informer} = \mathcal{D}(Encoder(X))
	\end{equation}
	where the decoder structure $\mathcal{D}$ is similar to the encoder structure $\mathcal{E}$ and consists of multiple $\mathcal{B_E}$. Here, the $\mathcal{B_E}$ formula is shown as follows:
	\begin{equation}
	\mathcal{B_E}(X) = f_{\text{conv}_2}(f_{\text{conv}_1}(\mathcal{A}(X)))
	\end{equation}
	where $f_{\text{conv}_2}$ and $f_{\text{conv}_1}$ are convolutional neural networks and $\mathcal{A}$ is ProbSparse self-attention.

	\subsubsection{Advantages of the Informer}
	
	Electric vehicle charging demand exhibits complex periodic patterns, including daily, weekly, and seasonal cycles, which often span hundreds or even thousands of time steps. Traditional sequence models struggle with vanishing gradients and high computational complexity when handling such long sequences \cite{cui2024informer}. The Informer model effectively overcomes these limitations by leveraging its ProbSparse attention mechanism \cite{zhang2024multimodal}. This approach selectively focuses on the most relevant key queries, significantly reducing computational overhead while preserving the ability to capture long-range dependencies. Unlike standard self-attention, which scales quadratically with sequence length, Informer’s sparse attention mechanism ensures efficient long-sequence processing. Additionally, the Informer model demonstrates strong adaptability across various time-series forecasting tasks, exhibiting high generalization capabilities even in data-scarce environments \cite{li2024deep}.

	\subsection{KAN Module}
	The KAN (Kolmogorov-Arnold Networks) module is a novel neural network architecture based on the Kolmogorov-Arnold representation theorem. It is specifically designed to enhance nonlinear function approximation by replacing traditional fixed activation functions with learnable activation functions, allowing for greater flexibility and improved model expressiveness.
	
	\begin{figure}[!t]
		\centering
		\includegraphics[scale=0.5]{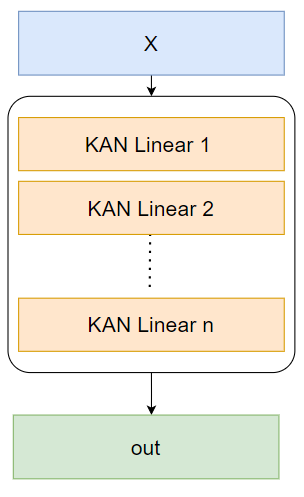}
		\caption{Overall architecture of the KAN}
	\end{figure}
	
	\subsubsection{Network Structure}
	The core innovation of the KAN module lies in replacing traditional fixed activation functions at the nodes (neurons) with learnable activation functions assigned to the edges (weights). This design enhances the model’s ability to approximate complex nonlinear functions. The network structure consists of the following key components:
	
	\begin{itemize}
		\item \textbf{Input Layer:} Receives input features.
	\end{itemize}
	\begin{itemize}
		\item \textbf{Edge Activation Function Layer:} Instead of using predefined activation functions at the nodes, learnable univariate activation functions are applied to each edge. These functions, which are often represented by splines or other parameterized functions, dynamically adapt during training to improve flexibility.
	\end{itemize}
	\begin{itemize}
		\item \textbf{Node Linear Combination Layer:} Each node aggregates the outputs from all its incoming edges using a linear combination, forming the final transformed representation.
	\end{itemize}
	Compared to traditional Multi-Layer Perceptrons (MLPs), KAN transfers nonlinearity from nodes to edges, thereby improving the model's flexibility. The specific calculation formula is shown as follows:
	\begin{equation}
	y = \sum f_{ij}(X_i)
	\end{equation}
	where $f_{ij}$ is the learnable activation function connecting input feature $x_i$ and output node $j$. Each $f_{ij}$ can be represented as follows:
	\begin{equation}
	f_{ij}(x_i) = \sum c_k * b_k(x_i)
	\end{equation}
	where $b_k$ are basis functions (such as B-spline basis functions) and $c_k$ are learnable coefficients.
	
	\subsubsection{Advantages of the KAN}
	The KAN module enhances function approximation by incorporating learnable activation functions, particularly spline-based functions \cite{li2025enhancing}. Unlike traditional neural networks with fixed activation functions, KAN dynamically adjusts nonlinear transformations, improving its adaptability and expressiveness. By modulating spline function control points, the model can efficiently approximate arbitrary nonlinear functions \cite{malik5123194improvised}. Additionally, the integration of sparsity constraints enhances both computational efficiency and interpretability, making the model more practical for real-world applications. In comparison to traditional MLPs, the KAN module demonstrates superior function approximation capabilities, particularly in high-dimensional data processing. Furthermore, its learnable activation functions enable deeper insights into the model’s behavior, providing a new perspective for interpretable neural network design \cite{espana2024estimating}.
	
	\subsection{Two-Stage Transfer Learning Strategy}
	
	To mitigate the data scarcity challenge in newly deployed EV charging stations, this study introduces a two-stage transfer learning framework designed to enhance both the accuracy and robustness of charging load forecasting. This framework leverages large-scale historical data from existing charging stations and optimizes model performance through a structured pre-training and fine-tuning process. By first learning generalizable patterns from established stations and then adapting to data-limited environments, this approach significantly improves forecasting performance in newly deployed stations.
	
	\subsubsection{Pre-Training}
	
	The pre-training stage utilizes large-scale historical data from operating EV charging stations as the source domain. These data contain extensive records of charging behaviors, enabling the model to learn the complex relationships between common time-series patterns and multidimensional external factors. To ensure effective training, the selection of an appropriate optimizer is crucial. Adam or AdamW is typically used due to their strong convergence properties and stability in time-series forecasting tasks. The learning rate is set at a moderate level (e.g., 0.001 to 0.0001) to balance training speed and prediction accuracy. To prevent overfitting on the source domain data, dropout regularization (dropout rate between 0.1 and 0.3) is applied, along with L2 weight decay (e.g., 0.0001) to improve generalization. The pre-training objective is to optimize model parameters by minimizing the Mean Squared Error (MSE) loss function, which is given by
	\begin{equation}
	L_{source} = \frac{1}{N}\sum (y_i - y'(\theta))^2
	\end{equation}
	where $y_i$ denotes the ground truth charging load, $y'(\theta)$ represents the model’s prediction, and $N$ denotes the total number of training samples.
	
	\subsubsection{Fine-Tuning and Transfer}
	
	The pre-trained parameters $\theta$ are transferred to newly deployed charging stations (target domain) with limited data. Since the pre-trained model has already learned general EV charging behavior patterns, this transfer process reduces dependence on extensive target domain data while enhancing initial prediction accuracy. Although newly constructed stations typically provide only short-term operational records, transfer learning enables the model to quickly adapt to the new environment. To further refine the model’s performance in the target domain, the following fine-tuning strategies are employed:
	\begin{itemize}
		\item \textbf{Layer Freezing:} The weights of the Mixer and Informer modules from the pre-trained model are retained, while only the KAN module is fine-tuned to capture station-specific characteristics, in order to ensure efficient adaptation with minimal data.
	\end{itemize}
	\begin{itemize}
		\item \textbf{Partial or Full Fine-Tuning:} All model parameters are updated using a reduced learning rate (e.g., one-tenth of the initial rate), with early stopping employed to prevent overfitting.
	\end{itemize}
	\begin{itemize}
		\item \textbf{Small-Batch Learning:} When training data is extremely limited, small batch sizes (e.g., 8 or 16) are used to optimize learning efficiency. A validation set is utilized to monitor generalization performance and prevent model degradation.
	\end{itemize}
	\noindent Fine-tuning is performed by minimizing the loss function on the target domain dataset, which can be expressed as
	\begin{equation}
	L_{target} = \frac{1}{M}\sum (y_i - y'(\theta))^2
	\end{equation}
	where $M$ is the number of samples in the target dataset.
	
	\section{EXPERIMENTS}
	
	To rigorously evaluate the performance of the MIK-TST, this study conducted a comprehensive experimental validation utilizing historical data collected from 26 charging stations in Boulder, USA\cite{boulder_ev_charging}. The experiments aim to address the following key research questions: 
	
	\begin{itemize}
		\item (RQ1): How does the performance of MIK-TST compare to other state-of-the-art forecasting models?
		\item (RQ2): What is the impact of different MIK-TST modules on overall prediction performance?
		\item (RQ3): How do different hyperparameter settings influence model performance?
	\end{itemize}
	
	\subsection{Experimental Setting}
	
To assess the effectiveness of MIK-TST, we conducted experiments using historical data from 26 EV charging stations in Boulder, USA. This dataset comprises detailed records of charging activities, including charging times, power consumption, user behavior patterns, and other station-level indicators. The dataset serves as a valuable resource for analyzing EV charging behaviors, forecasting charging demand, and optimizing charging station operations. Its rich temporal and user-related attributes make it particularly well-suited for evaluating the performance of advanced time-series forecasting models in real-world scenarios.

To design a realistic transfer learning setting, we first assigned sequential indices (0 to 25) to the 26 stations based on the chronological order of their first recorded charging activity. Stations indexed from 0 to 20 were selected as the source domain and used for pre-training the model with their complete historical charging data. Subsequently, the pre-trained model was fine-tuned using a small amount of early data (prior to 2023) from target stations indexed 21 to 25. After fine-tuning, the model was evaluated on its ability to predict future charging loads at these newly deployed target stations, simulating real-world application scenarios where only limited early data is available for adaptation.

	\subsection{Methods for Comparison}
	
	We compared MIK-TST with various baseline methods from different research lines as follows:
	
	\textbf{PatchTST}\cite{nie2022time}: achieves efficient and accurate long-sequence time-series forecasting by segmenting time-series data into non-overlapping patches and processing them with a Transformer-based architecture.
	
	\textbf{Autoformer}\cite{wu2021autoformer}: enhances forecasting accuracy by introducing a decomposition module and an auto-correlation mechanism to effectively capture periodic patterns and long-range dependencies in time-series data.
	
	\textbf{DLinear}\cite{chen2022transformers}: uses a stacked linear layer architecture combined with a decomposition module to efficiently model both trend and seasonal components.
	
	\textbf{Crossformer}\cite{zhang2023crossformer}: enhances the accuracy of multivariate time-series forecasting by incorporating a cross-dimension attention mechanism to effectively capture complex interactions between different dimensions of time-series data.
	
	\textbf{FreTS}\cite{yi2023frequency}: focuses on frequency-domain learning to extract key spectral features, effectively reducing noise interference and improving computational efficiency through a streamlined deep learning pipeline.
	
	\subsection{Parameter Settings}
	
	We implemented MIK-TST using PyTorch and set the embedding size of all modules to 256. Embeddings were initialized using the Normal and Xavier initialization methods. For optimization, we used the Adam optimizer with a learning rate of 1e-4 and weight decay set to 0. The multi-layer neural network in the model was configured with a layer structure of [256, 2048, 256], and the maximum number of iterations was set to 10. For the baseline models, we followed the original parameter settings provided in their respective papers to ensure a fair performance comparison.
	
	\subsection{Performance Validation (RQ1)}
	
	To assess the effectiveness of our proposed MIK-TST, we evaluate its performance using Mean Absolute Error (MAE) and Mean Squared Error (MSE), which are two widely used metrics balancing prediction accuracy and robustness. MAE measures the average magnitude of errors in predictions, while MSE emphasizes larger errors by squaring the differences, providing a robust indicator of forecasting precision. These metrics are defined as follows:
    
    \textbf{Mean Absolute Error (MAE):}
    \begin{equation}
    \text{MAE} = \frac{1}{N} \sum_{i=1}^{N} |y_i - \hat{y}_i|
    \end{equation}

    \textbf{Mean Squared Error (MSE):}
    \begin{equation}
    \text{MSE} = \frac{1}{N} \sum_{i=1}^{N} (y_i - \hat{y}_i)^2
    \end{equation}

where \( y_i \) denotes the ground truth charging load, \( \hat{y}_i \) represents the model's prediction, and \( N \) is the total number of samples.
	
	\begin{table}[!t] 
		\centering 
		\caption{Performance comparison on different models} 
		\begin{tabular}{lcccccccc} 
			\toprule 
			Model & MAE & MSE  \\
			\midrule
			PatchTST & 0.4638 & 0.7409 \\
			Autoformer & 0.4845 & 0.6869\\
			DLinear & 0.4751 & 0.7293\\
			Crossformer & 0.4726 & 0.7319 \\
			FreTS & 0.4670 & 0.7282 \\
			MIK-TST (OURS) & \textbf{0.4450(4\%)} & \textbf{0.6322 (8\%)} \\
			
			\bottomrule 
		\end{tabular}
		
		\label{tab:performance} 
	\end{table}

	Table~\ref{tab:performance} presents a comprehensive comparison of MIK-TST against state-of-the-art forecasting models. MIK-TST achieves an MAE of 0.4450 and an MSE of 0.6322, outperforming the best baseline, PatchTST (MAE: 0.4638, MSE: 0.7409), by 4\% in MAE and 8\% in MSE. This improvement stems from MIK-TST's integrated design: the Mixer module enhances prediction accuracy by effectively fusing temporal and channel features, reducing errors in capturing multi-dimensional charging patterns (e.g., daily peaks and user-specific trends). The Informer module, with its ProbSparse attention, excels at modeling long-range dependencies---critical for EV charging data spanning weeks or months---resulting in a lower MSE by preserving temporal consistency. The KAN module's learnable activation functions further refine nonlinear approximations, enabling the model to adapt to complex, station-specific load variations, as evidenced by its superior performance in data-scarce target stations.
    
    In contrast, PatchTST, while efficient for long-sequence forecasting via patching, lacks a robust feature fusion mechanism, leading to higher MAE due to incomplete utilization of multi-source data. Autoformer captures periodic patterns effectively but struggles with computational scalability for extended horizons, reflected in its higher MSE (0.6869). DLinear’s linear architecture, though lightweight, fails to model the nonlinear dynamics of EV charging (MSE: 0.7293), while Crossformer and FreTS, despite their strengths in cross-dimensional and frequency-domain modeling, exhibit limited adaptability to new stations without transfer learning, as seen in their MAE values (0.4726 and 0.4670). For newly constructed stations, where accurate forecasting can reduce energy waste by up to 8\% (based on MSE gains), MIK-TST’s two-stage transfer learning ensures rapid adaptation, making it a practical solution for smart grid optimization.
	
	\subsection{Model Ablation Study (RQ2)}
		
\begin{table}[h]
    \centering
    \caption{Ablation Experiment Results}
    \label{tab:ablation}
    \begin{threeparttable}
        \begin{tabular}{lcc}
            \toprule
            Model & MAE & MSE \\
            \midrule
            w/o-Mcl & 0.4582 & 0.6540 \\
            w/o-Kcl & 0.4508 & 0.6338 \\
            w/o-Dcl & 0.4772 & 0.7457 \\
            MIK-TST & \textbf{0.4450} & \textbf{0.6322} \\
            \bottomrule
        \end{tabular}
    \end{threeparttable}
    \smallskip \\
    \small \raggedright \textit{Remark:}\\
    \small \raggedright Where “w/o” denotes the exclusion of a specific component. \\ 
    “w/o-Mcl”, “w/o-Kcl”, and “w/o-Dcl” represent the removal of the Mixer module, the KAN module, and the two-stage transfer learning strategy, respectively.
\end{table}
	
	To evaluate the contribution of each MIK-TST component, we conducted ablation studies by systematically excluding the Mixer (w/o-Mcl), KAN (w/o-Kcl), and two-stage transfer learning (w/o-Dcl) modules, as shown in Table~\ref{tab:ablation}, where ``w/o'' denotes the exclusion of a specific component. Removing the Mixer module (w/o-Mcl) increases MAE from 0.4450 to 0.4582 and MSE from 0.6322 to 0.6540, a 3\% and 3.5\% degradation, respectively. This underscores Mixer’s role in fusing multi-source features, which is critical for capturing interactions between temporal patterns and station-specific indicators (e.g., day-of-week effects). Without it, the model struggles to integrate diverse inputs, leading to higher errors in peak load predictions. Excluding the KAN module (w/o-Kcl) raises MAE to 0.4508 and MSE to 0.6338, indicating a modest 1.3\% MAE increase but minimal MSE impact. This suggests KAN’s learnable activation functions enhance fine-grained nonlinear modeling, though its absence is less detrimental in stations with simpler load patterns. The most pronounced effect occurs when removing the transfer learning strategy (w/o-Dcl), with MAE rising to 0.4772 (7.2\% increase) and MSE to 0.7457 (18\% increase). This highlights transfer learning’s indispensable role in adapting pre-trained knowledge to data-scarce stations, a common challenge in real-world deployments where early operational data is limited to weeks rather than years.
    
    These results align with MIK-TST’s design goals: Mixer ensures robust feature representation, KAN refines complex dynamics, and transfer learning bridges data gaps. Notably, the interplay between Mixer and transfer learning appears synergistic---without Mixer’s rich features, pre-training on source stations loses effectiveness, amplifying errors in target stations. This validates MIK-TST’s holistic approach for EV charging forecasting.

		\begin{figure*}[!t]
		\centering
		\subfloat{\includegraphics[scale=0.2]{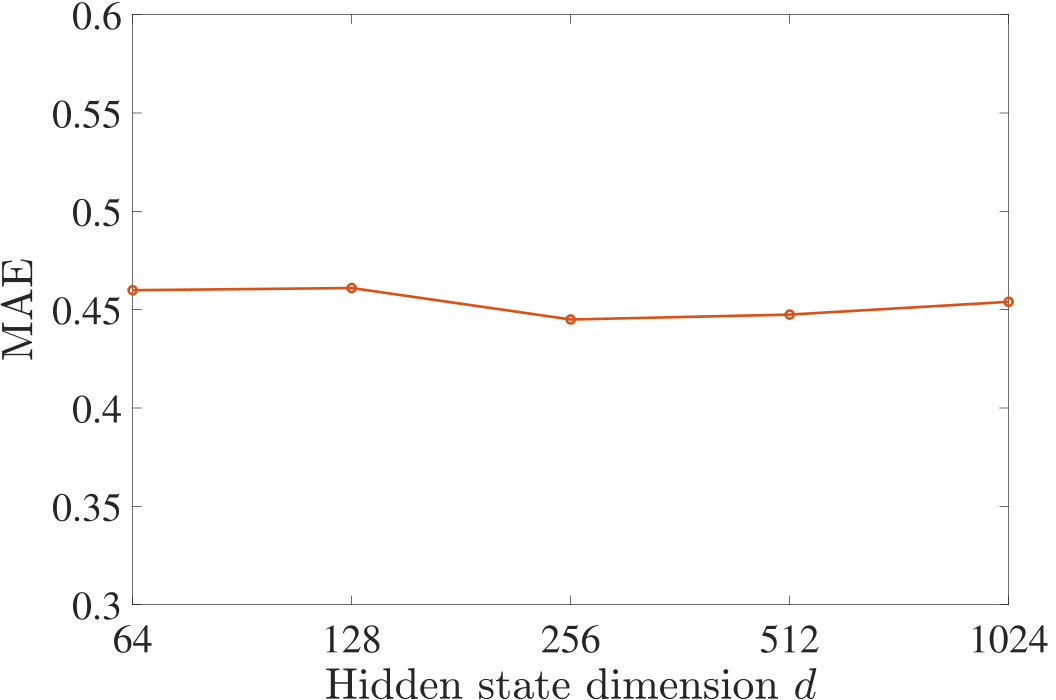}}  \hspace{1.5cm}
		\subfloat{\includegraphics[scale=0.2]{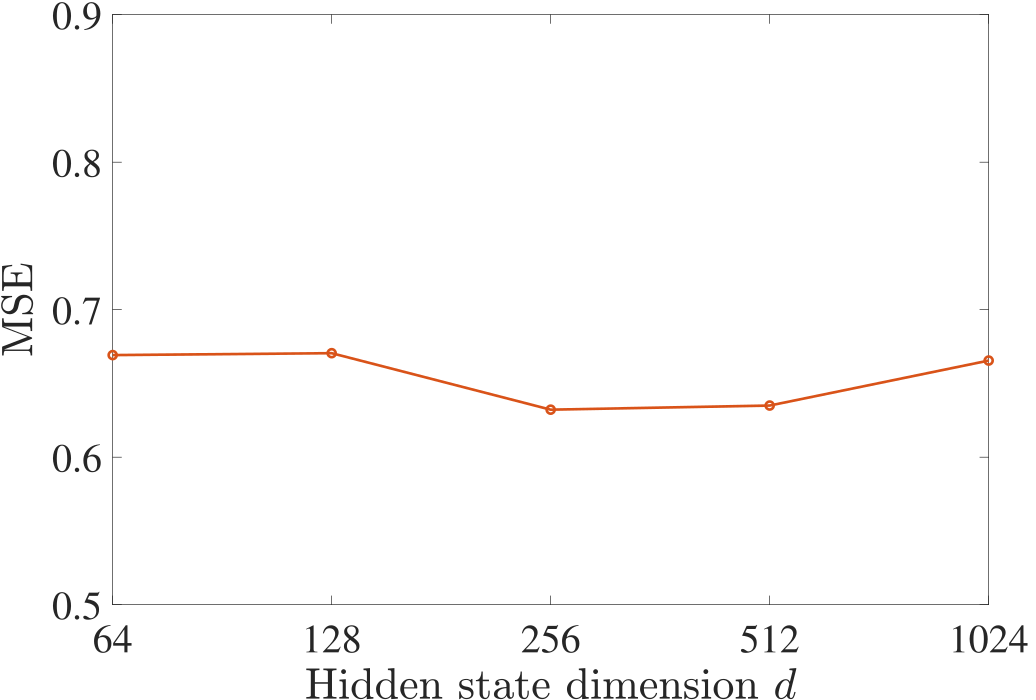}}\hspace{0.7cm}
		\caption{Parameter sensitivity analysis of MIK-TST with respect to the hidden state dimension $d$.}
		\label{RT3_1}
	\end{figure*}
	
	\begin{figure*}[!t]
		\centering
		\subfloat{\includegraphics[scale=0.2]{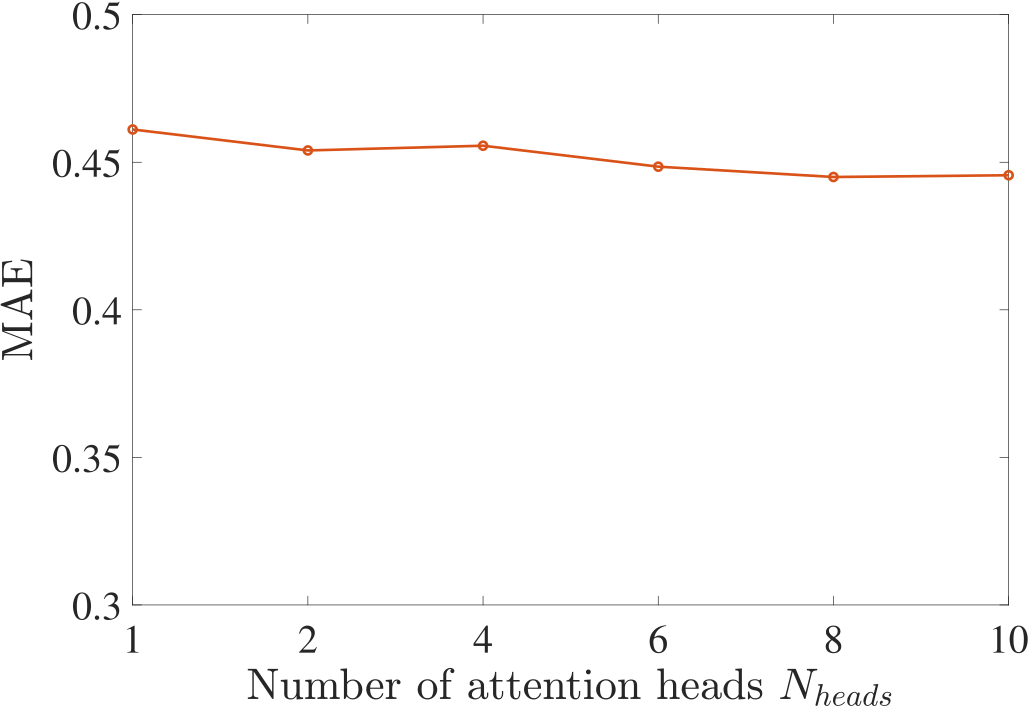}}  \hspace{1.5cm}
		\subfloat{\includegraphics[scale=0.2]{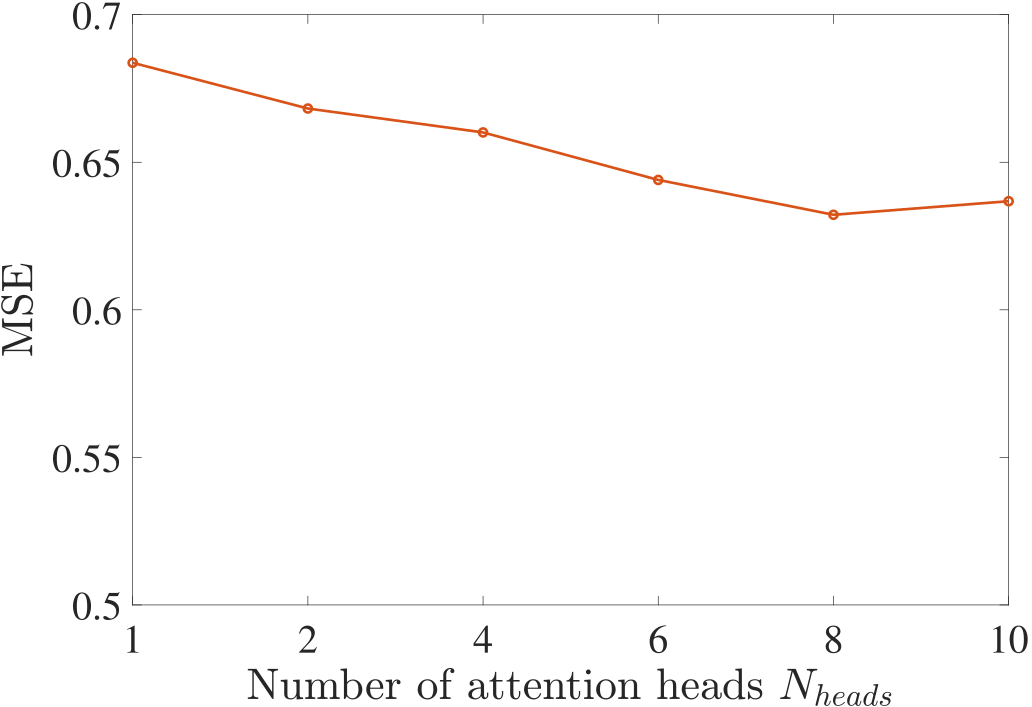}}\hspace{0.7cm}
		\caption{Parameter sensitivity analysis of MIK-TST with respect to the number of attention heads $N_{heads}$.}
		\label{RT3_2}
	\end{figure*}
	
	\begin{figure*}[!t]
		\centering
		\subfloat{\includegraphics[scale=0.2]{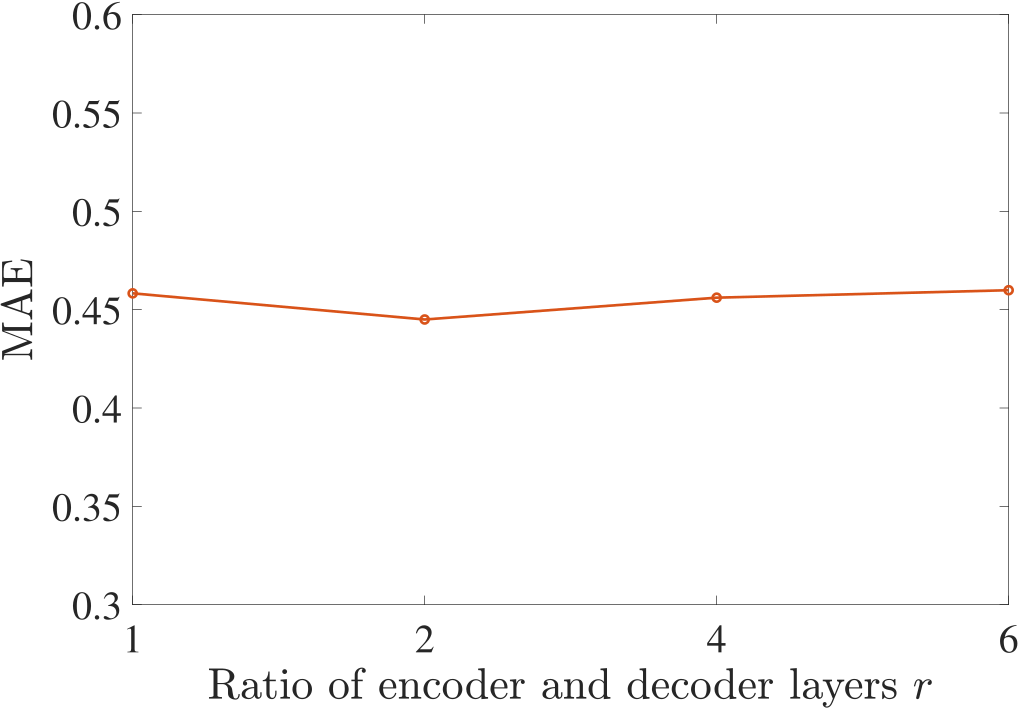}}  \hspace{1.5cm}
		\subfloat{\includegraphics[scale=0.2]{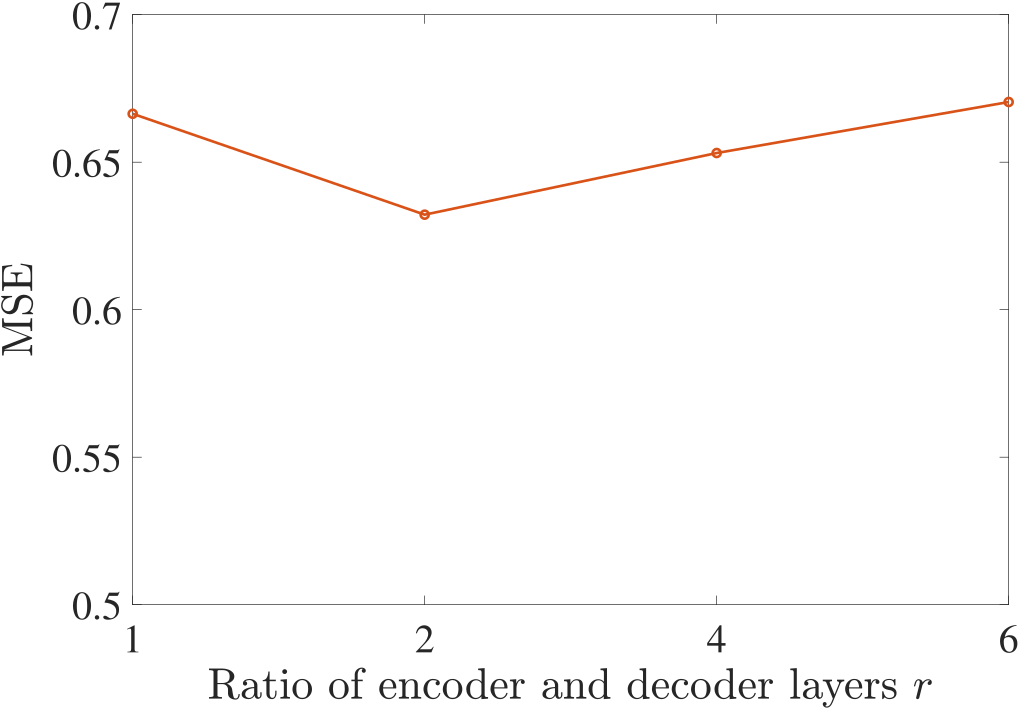}}\hspace{0.7cm}
		\caption{Parameter sensitivity analysis of MIK-TST with respect to the number of encoder and decoder layers $r$}
		\label{RT3_3}
	\end{figure*}

	\subsection{Hyperparameter Analyses (RQ3)}

	To investigate the sensitivity of MIK-TST to key hyperparameters, we systematically varied the hidden state dimension \(d\), the number of attention heads \(N_{\text{heads}}\), and the number of encoder/decoder layers \(r\), as depicted in Figs.~\ref{RT3_1}-\ref{RT3_3}. These parameters directly influence the model's capacity, attention mechanism, and depth, respectively, playing critical roles in balancing representational power, computational efficiency, and generalization for long-sequence EV charging load forecasting. Below, we analyze their effects in detail, supported by quantitative results and their implications within the MIK-TST architecture.

\textbf{Hidden State Dimension (\(d\))}: We tested \(d\) across \{64, 128, 256, 512, 1024\}, which determines the embedding size of feature representations across the Mixer, Informer, and KAN modules. At \(d=64\), the model yields an MAE of 0.4599 and MSE of 0.6692, reflecting limited capacity to encode the high-dimensional, multi-source features (e.g., temporal patterns, charging power, session logs) typical of EV data. Increasing \(d\) to 256 reduces MAE to 0.4450 and MSE to 0.6322, as the larger embedding space enhances the Mixer’s ability to fuse heterogeneous inputs and the KAN’s capacity to model nonlinear relationships. This aligns with the need to capture complex dependencies, such as daily/weekly charging cycles, in long-sequence data. However, at \(d=512\), MAE rises to 0.4475 and MSE to 0.6350, and at \(d=1024\), MAE further increases to 0.4540, indicating overfitting. This occurs because excessively large embeddings introduce redundant parameters, overfitting the pre-training data and reducing adaptability during fine-tuning on sparse target station data. Thus, \(d=256\) strikes an optimal balance, providing sufficient expressiveness without compromising generalization.

\textbf{Number of Attention Heads (\(N_{\text{heads}}\))}: We evaluated \(N_{\text{heads}}\) in \{1, 2, 4, 6, 8, 10\}, a parameter specific to the Informer module’s ProbSparse attention mechanism, which governs how the model attends to different temporal dependencies. With a single head (\(N_{\text{heads}}=1\)), MAE is 0.4611 and MSE is 0.6837, as the model focuses on a narrow subset of dependencies, missing diverse patterns like peak-hour surges or weekend trends. Performance improves with more heads, peaking at \(N_{\text{heads}}=8\) (MAE: 0.4450, MSE: 0.6322), where multiple heads enable the Informer to capture varied long-range relationships---e.g., correlating current loads with historical peaks days or weeks prior. This is critical for EV charging, where periodicities span multiple timescales. Beyond this, at \(N_{\text{heads}}=10\), MAE rises to 0.4456 and MSE to 0.6368, due to increased computational noise and attention dilution, where redundant heads focus on less relevant features, degrading efficiency. Hence, \(N_{\text{heads}}=8\) optimizes the trade-off between capturing diversity and maintaining focus in attention.

\textbf{Number of Encoder and Decoder Layers (\(r\))}: We varied \(r\) (i.e., \(e_{\text{layers}}/d_{\text{layers}}\)) across \{1, 2, 4, 6\}, controlling the depth of the Informer’s encoder-decoder structure. With \(r=1\), MAE is 0.4583 and MSE is 0.6664, indicating insufficient depth to extract hierarchical temporal features from long sequences, limiting the model’s ability to model trends across weeks. At \(r=2\), performance peaks (MAE: 0.4450, MSE: 0.6322), as the additional layer allows the encoder to refine high-level representations (e.g., seasonal patterns) and the decoder to generate precise long-horizon forecasts. Deeper configurations, e.g., \(r=4\) (MAE: 0.4561, MSE: 0.6531) and \(r=6\) (MAE: 0.4599, MSE: 0.6704), show declining performance due to overfitting, particularly on the limited fine-tuning data from new stations. Deep layers amplify small noise in sparse datasets, reducing generalization. Thus, \(r=2\) provides an effective depth for EV forecasting, balancing complexity and robustness.

These analyses reveal that \(d=256\), \(N_{\text{heads}}=8\), and \(r=2\) optimize MIK-TST’s performance by aligning with its architectural strengths: Mixer’s feature fusion benefits from moderate embedding sizes, Informer’s attention thrives with multi-head diversity, and the encoder-decoder depth supports hierarchical modeling without overfitting. For EV charging stations, where long-sequence data and data scarcity coexist, these settings ensure accurate, adaptable predictions, as validated by Figs.~\ref{RT3_1}-\ref{RT3_3}.

	\section{CONCLUSION}

	In this work, we introduce MIK-TST, a novel two-stage transfer learning framework that synergistically integrates the Mixer, Informer, and Kolmogorov-Arnold Networks (KAN) architectures to address the critical challenge of electric vehicle (EV) charging load forecasting, particularly in newly constructed charging stations. The Mixer module excels at fusing multi-source heterogeneous features across temporal and channel dimensions, enabling the model to leverage comprehensive input data effectively. The Informer module, with its ProbSparse attention mechanism, adeptly captures long-range temporal dependencies and periodic patterns inherent in charging load time series, overcoming the limitations of traditional models in handling extended sequences. Meanwhile, the KAN module enhances the framework’s capability to model complex nonlinear relationships through its innovative use of learnable activation functions, simultaneously improving both predictive accuracy and interpretability. By incorporating a two-stage transfer learning strategy—pre-training on large-scale data from established stations followed by fine-tuning on limited data from new stations—MIK-TST effectively mitigates the pervasive issue of data scarcity, ensuring robust generalization and adaptability in real-world scenarios. 
    
    Our experimental validation, conducted on a dataset comprising historical charging records from 26 stations in Boulder, USA, demonstrates that MIK-TST outperforms state-of-the-art baseline models, achieving improvements of 4\% in Mean Absolute Error (MAE) and 8\% in Mean Squared Error (MSE). These gains underscore the framework’s superior ability to integrate diverse features, model long-term dependencies, and adapt to data-scarce environments. Beyond its technical contributions, MIK-TST offers significant practical value for smart grid management and urban energy planning. By providing accurate and scalable load forecasts, it supports efficient energy distribution, optimizes resource allocation, and facilitates the sustainable expansion of EV infrastructure—key pillars of smart city development and the global transition to clean energy.
    
    Despite its strengths, MIK-TST is not without limitations. The current framework primarily relies on station-specific historical data and does not fully account for external variables, such as weather conditions, traffic patterns, or socio-economic factors, which could further influence EV charging demand. Additionally, while the two-stage transfer learning approach enhances adaptability, its performance in highly heterogeneous or rapidly evolving environments (e.g., stations with drastically different usage patterns) warrants further investigation. Computational efficiency, though improved by the ProbSparse attention mechanism, could also pose challenges for real-time deployment on resource-constrained systems. 
    
    Looking ahead, several avenues for future research emerge. First, enhancing domain adaptation techniques—such as incorporating adversarial training or domain-specific regularization—could improve the model’s robustness across diverse charging station profiles. Second, integrating external factors, such as real-time weather data, traffic flow, or user demographics, into the feature set could yield more comprehensive and context-aware predictions. Third, optimizing the computational efficiency of MIK-TST, potentially through model pruning or quantization, would facilitate its deployment in real-time energy management systems. Finally, incorporating uncertainty estimation techniques, such as Bayesian methods or ensemble approaches, could further enhance the model’s reliability by quantifying prediction confidence, a critical feature for decision-making in intelligent energy systems. By addressing these opportunities, future iterations of MIK-TST could provide even greater accuracy, scalability, and practical utility, solidifying its role as a cornerstone for next-generation EV charging infrastructure management.

\end{document}